# Non-thermal Magnetic Deicing Using Two-Dimensional Chromium Telluride

*Chinmayee Chowde Gowda[1], Alexey Kartsev[2,3,4*], Nishant Tiwari[5], Safronov A.A[3], Prafull Pandey[6], Ajit K. Roy[7], Pulickel M. Ajayan[8], Douglas S. Galvão[9*] and Chandra Sekhar Tiwary[1,5]*

[1]School of Nano Science and Technology, Indian Institute of Technology Kharagpur, West Bengal – 721302, India

[2]Computing Center of the Far Eastern Branch of the Russian Academy of Sciences, 680000 Khabarovsk, Russia

[3]MIREA-Russian Technological University, 119454 Moscow, Russia

[4]Peoples' Friendship University of Russia (RUDN University), 6 Miklukho-Maklaya St, Moscow 117198, Russia

[5]Department of Metallurgical and Materials Engineering, Indian Institute of Technology Kharagpur, West Bengal – 721302, India

[6]Materials Engineering, Indian Institute of Technology Gandhinagar, Gujarat – 382055, India

[7]Materials and Manufacturing Directorate, Air Force Research Laboratory, Wright Patterson AFB, Ohio, 45433-7718, USA

[8]Department of Materials Science and NanoEngineering, Rice University, 6100 S Main Street, Houston, TX 77005, USA

[9]Applied Physics Department and Computational Engineering and Sciences, State University of Campinas, Campinas, SP, 13083-970, Brazil

SYNOPSIS: Icing of surfaces is a major issue which effects the livelihood of people residing in northern hemisphere. The study focuses on deicing of surfaces with ferromagnets near room temperature.

ABSTRACT

Two-dimensional (2D) chromium telluride $Cr_2Te_3$ exhibits strong ferromagnetic ordering with high coercivity at low temperatures and paramagnetic behavior when approaching room temperature. The spin states of monolayer $Cr_2Te_3$ shows ferromagnetic ordering in the ground state, and *in-situ* Raman analysis shows reversible structure transformation and hence ferromagnetic to paramagnetic transition during low-temperature heating cycles (0 - 25 °C). The magnetic phase transition near room temperature in the 2D $Cr_2Te_3$ prompted the study of these layered material exploration for energy application. We demonstrate that the low-temperature ferromagnetic behavior can be used to magnetically deice material surfaces using an external magnetic source, avoiding the use of harsh chemicals and high temperatures. The hydrophobic nature and dipole interactions of $H_2O$ molecules with the surface of the 2D $Cr_2Te_3$ coating aid in the condensation of ice droplets formed on the surface. First-principles calculations also confirm the observed crystal structure, surface interaction, and magnetic properties of 2D $Cr_2Te_3$.

INTRODUCTION

Chalcogenides with layered and non-layered structures present versatile magnetic behavior when their dimensions are reduced. Investigations in the field of room temperature (RT) two - dimensional (2D) magnets started with transition metal di-chalcogenides (TMDCs) that exhibit thickness-dependent magnetic behavior. Among them, van der Waals magnets have become materials of interest [1,2] for their ease of cleaving into thin layers and their ability to produce 2D structures. TMDCs such as $Fe_3GeTe_2$ [3], $MnSe_x$ [4], $NbSe_2$ [5], $PtSe_2$ [6], $V_5Se_8$ [7], $VTe_2$ [8], $MoS_2$ [9], $ReS_2$, and Re-doped WSe$_2$ [10–12] were studied for long-range magnetic ordering effects at critical transition temperatures.

These materials pave way for spintronic devices and data storage units that can operate at ambient conditions. Although the Merwin-Wagner theorem forbade the presence of magnetic ordering in 2D isotropic Heisenberg systems with short-range interactions, we still observe magnetic ordering in the 2D atomic layer in lower dimensional Cr$_2$Te$_3$ structures. This is due to magneto-crystalline anisotropy in the 2D material. Chromium telluride (Cr$_2$Te$_3$) has a ferromagnetic ordering at RT and varying Curie temperatures ($T_C$) ranging from 180 K to 340 K [13,14], the behavior is also observed in simulations of pristine structures [15,16]. Strain engineering of 2D materials has the degree of freedom to alter the atomic thickness and the nature of the chemical bonds among layers, making the 2D materials mechanically flexible. The strain tunability is part of intrinsic magnetism. $T_C$ shifts have been observed upon applying tensile strain along with Berry curvature [16].

Magnetic particles have gained much attention in the design of functional surfaces and find potential applications in oil/water separation [17,18], hydrocarbon elimination [19], self-cleaning [20], and anti-icing [21,22]. Anti-icing/deicing is the process of clearing snow, slush, or ice from the fuselage, wings, and control surfaces. There are a number of standard techniques, including the use of heat, chemicals, rubber wipers, magnetic vibrations, and electromagnetic coils [23–25]. The use of chemical sprays and/or salts to melt this ice can be toxic at times. Propylene glycol is a liquid that is frequently heated to 140 °F and 150 °F and sprayed under pressure in chemical applications to eliminate ice and other impurities [26]. Other techniques involve electromagnetic coils with driver circuits[23], which produce heat and vibrations with a very high current flow. These methods, currently in use, often require considerable energy or involve the use of hazardous chemicals in deicing and anti-icing processes. New technologies including acoustic energy [27] usage and other renewable systems [28] have also expedite to synergistically defrost surfaces with reduced energy usage.

In this work, we obtained bulk $Cr_2Te_3$ by induction melting. Later, the samples were subjected to probe sonication to obtain 2D $Cr_2Te_3$ for large-scale synthesis. Strain was introduced in the structure via external stimulus during synthesis of 2D $Cr_2Te_3$. The ultrathin nature of the samples was confirmed by extensive structural and morphological characterizations. The crystal structure was simulated using density functional theory (DFT) methods, and the Cr1 layer was observed to be a magnetically active layer, which was used for further investigating the magnetic behavior. Herein, we report the paramagnetic nature of the 2D material at room temperature (300 K). A strong ferromagnetic behavior was observed at lower temperatures from < 250 K to 10 K. The obtained 2D $Cr_2Te_3$ structures were used as a coating to prevent icing at lower temperatures. We calculate the formation energy and van der Waal's (vdW) gap for the surface interaction between the $H_2O$ molecules and monolayer $Cr_2Te_3$. The lattice structure transitions were extensively studied using *in situ* Raman spectroscopy along with the wettability of the material coating on applications in ice-cold surface regions for deicing.

MATERIALS AND METHODS

EXPERIMENTAL DETAILS

Chromium (Cr) blocks (99 wt.%) and tellurium (Te) pieces (99 wt.%) were used as starting elements. Alloys with a sample weight of 15 g were prepared by induction melting under a high-purity argon (Ar) atmosphere. As there is a large difference between the melting points of Cr and Te, prior to melting, both elements are sealed in a quartz tube with backfilling of Ar to prevent the evaporation of Te. The samples were induction melted at 1200 °C and kept at a high temperature for 2 h before being brought down. The samples were then kept for heat treatment at 900 °C for 100 h to obtain homogeneous samples. The composition of the as-cast alloys was confirmed using a PAN Analytics X-ray fluorescence (XRF) instrument.

The exfoliated flakes were obtained via a liquid phase exfoliation route using isopropyl alcohol (IPA) as the solvent medium. A probe sonicator was used at a 30 kHz frequency for 8 h, with a pulse rate of 10 s. The suspension was allowed to rest for 24 h, and the supernatant was collected for further use.

CHARACTERIZATION

X-ray diffraction patterns and information on crystalline phases were analyzed by a Bruker D8 Advance X-ray diffractometer. The Cu-Kα source, operating at 40 kV voltage and 40 mA current, had a wavelength of (λ) 1.5406 Å. The Panalytical Xpert high score plus software and Pearson's Crystal Database were both utilized to analyze the XRD structural data. With a probe current of up to 20 nA, scanning electron microscopy (SEM) was carried out using an SEM CARL ZEISS SUPRA 40. We also used an electron probe microanalyzer (JEOL, JXA8530F) for the energy-dispersive X-ray spectroscopy analysis of the samples. Transmission electron microscopy (TEM) was used to analyze the samples further. TEM imaging was performed in 60 – 300 kV low base Titan® Themis™ with a monochromator, a CEOS probe corrector for Cs aberration-corrected STEM, and a Quantum 965 Gatan Imaging Filter (post column) for EELS analysis. The exfoliated samples were dispersed with ethanol and sonicated for 10 mins. Later, 1-2 drops of the dispersed sample were drop cast on the TEM grid. The TEM grid was a carbon-coated copper grid. Specification: Square mesh 500. The HAADF – STEM images were obtained under 300 kV voltage. Atomic force microscopy (AFM) was carried out utilizing an Agilent Technologies Model No. 5500. On a monocrystalline silicon substrate, the sample was deposited using the drop deposition technique. Raman spectroscopy analysis was carried out using a WITec UHTS Raman spectrometer (WITec, UHTS 300 VIS, Germany) with a laser excitation wavelength of 532 nm at room temperature (RT). To investigate the oxidation states and surface composition of the materials, an XPS ThermoFisher Scientific Nexsa was used, with Al–Kα radiation ($\lambda=1486.71$ eV) as a source. Magnetic measurements were performed using a Quantum Design MPMS SQUID VSM EverCool system with an operating temperature range from 1.8 K to 1000 K.

DFT SIMULATIONS

We carried out density functional theory (DFT) simulations to study bulk and 2D $Cr_2Te_3$ crystals in the frameworks of VASP package[29] and projector augmented wave (PAW) formalism[30]. Structural relaxation procedures were performed at ambient pressure based on the total energy and forces minimization principle [31,32], with convergence criteria of $10^{-8}$ eV and $10^{-7}$ eV/Å, respectively. Spin-polarized formulation of DFT and GGA+U formalism[33] coupled to spin-orbit coupling (SOC) corrections were used to calculate the non-collinear states of magnetic subsystems. Convergence of results against numerical parameters was tested, and the energy cutoff 350 eV along with the 8×8×8 and 8×8×1 k-point grid was employed for bulk and monolayered structures, respectively. The final 2D structure of $Cr_2Te_3$ is shown in **Figure S5a**. **Figure S5b** shows the magnetic sublattice of chromium. The on-site Hubbard correction for Cr *d*-orbitals in the framework of Dudarev approach[34] was chosen *U*=2.65 eV, as mean *U* values previously employed for chromium selenites and tellurites[35–37]. The unit cell preparation and structural visualization were done by employing VESTA program[38].

To address the wetting/non-wetting nature of the paramagnetic/ferromagnetic 1L-$Cr_2Te_3$, we have utilized 2×2 1L-$Cr_2Te_3$ and 3×3 $H_2O$ supercells. Two cases of their heterostructure were considered both in ferromagnetic and paramagnetic states: oxygen and hydrogen ions oriented towards the chromium telluride plane (O- and H-oriented). The k-point mesh was chosen to be 4×4×1, and van der Waals interactions were accounted for by means of the empirical D3 correction (BJ) correction from Grimme et al. [39]. The pseudo vacuum space between the periodic mappings along the z-axis was chosen to be 20 Å to avoid spurious interactions of mirrored layers.

RESULTS

Induction melted chromium telluride ($Cr_2Te_3$) samples were crushed manually and then sonicated for eight hours to obtain ultrathin flakes of $Cr_2Te_3$ suspended in IPA solvent (shown in **Figure S1a**). The phase formation of the samples was confirmed through X-ray diffraction (XRD). From **Figure 1a,** we confirmed pure phase formation of bulk $Cr_2Te_3$ with a hexagonal Ni-As structure ($P\bar{3}1c$ (163) space group) with ordered Cr vacancies. Chromium chalcogenide structure exhibits a distorted hexagonal close packing of Te atoms and Cr atoms in octahedral interstices[40]. The unit cell consists of alternating Cr and Te layers, with Cr vacancies in every second metal layer. There are four vertically aligned Cr sites, with one layer being weakly antiferromagnetic in nature. The hexagonal *c*-axis is the magnetic access with perpendicular magnetic field alignment. The diffraction peaks are indexed with the (110), (102), (103), and (201) planes of the hexagonal $Cr_2Te_3$ phase (JCPDS card no. 01-071-2246). The exfoliated 2D $Cr_2Te_3$ also had pure phase formation with no oxide peaks, and (201) plane exfoliated in excess. The lattice parameters were calculated to be $a = b = 6.82$ Å and $c = 11.92$ Å. Due to weak interlayer coupling the corresponding structural model was created to eliminate all other interactions except van der Waals forces. This model was used for the subsequent DFT electrical and magnetic calculations.

In **Figure S2**, we present a volumetric view of the original bulk structure along with the obtained initial 2D structure. The initial $Cr_2Te_3$ structural configuration was taken from the Electronic Resource Material Project. **Figure S2a** and **S2b** show a three-dimensional and polyhedral view of the structure of the material (hexagonal lattice); Cr and Te atoms are shown in blue and in yellow, respectively. To model the 2D structure, the plane (1T0) was chosen (**Figure S2c**), after which the distance between the formed layers was increased to 25 Å (to prevent spurious interlayer interactions), then we obtained the initial 2D structure (**Figure S2d**). In this way, as mentioned above, we eliminated all types of interactions among the layers, except for the van der Waals ones, and then proceeded to geometry (relaxation) optimization.

Ultrathin flake formation was confirmed by AFM analysis (**Figure 1c**), where we observed a wide range of flake distributions with lateral flake sizes ranging from 2 to 8 nm, as seen in the inset of **Figure 1b**. The morphology of the samples was further analyzed by scanning electron microscopy (SEM), and elemental analysis (EDS) of bulk $Cr_2Te_3$ after homogenization for 48 h and 2D $Cr_2Te_3$ (**Figure S3 (a,b)**). **Figure S3b** shows exfoliated 2D $Cr_2Te_3$ composition with Cr: 20.68 wt% and Te: 79.32 wt%, as confirmed by EDS analysis. The chemical integrity of the exfoliated samples was verified using X-ray photoelectron spectroscopy (XPS). The XPS analysis of the exfoliated 2D $Cr_2Te_3$ sample has peaks of both Cr and Te. **Figure S3c** shows Cr2p with peaks observed at 577 eV ($2p_{3/2}$) and 586 eV ($2p_{1/2}$), with FWHMs of 3.1 and 3.5, respectively, along with satellite peaks at 574 and 584 eV. Similarly, Te3d peaks were observed at 577.35 for $3d_{5/2}$, which overlaps with the $Cr2p_{3/2}$ peak and 587.73 eV for $3d_{1/2}$ in **Figure S3d**. A slight redshift of ~ 0.5 eV was observed in the exfoliated samples compared to their bulk counterpart, which is generally attributed to the strain in the exfoliated structure induced during probe sonication.

The high-angle annular dark-field scanning transmission electron microscopy (HAADF-STEM) image of the exfoliated 2D $Cr_2Te_3$ flake was further analyzed to determine the atomic arrangement of Cr and Te atoms. Two different domain orientations were observed from the atomic arrangements, which were both hexagonal crystal systems. The bright dots correspond to tellurium (Te) atoms. Chromium (Cr) ones are not visible because of their smaller atomic radius. Cr atoms form intercepts between Te atoms. The FFT pattern was extracted from the HAADF STEM image (inset of **Figure 1c**), which shows the hexagonal crystal lattice of the exfoliated samples. **Figure 1c** shows an inverse FFT mapped from the same flake, showing a lattice spacing of 0.206 nm corresponding to the (1T0) plane, which was also further exfoliated during LPE. This value is in good agreement with the theoretically predicted lattice spacing planes $d = \frac{a_0}{2\sqrt{3}} \approx 0.195$ nm.

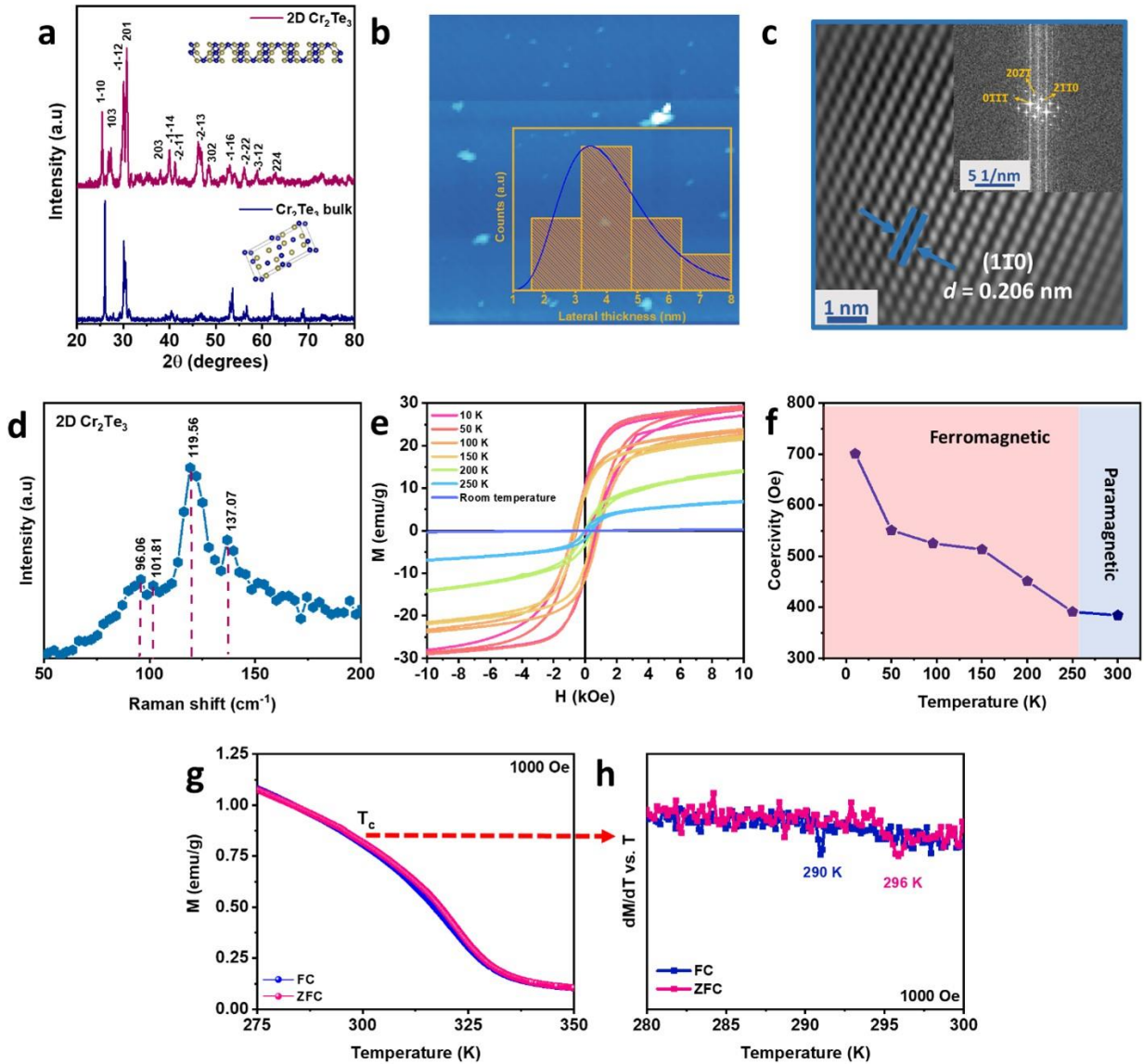

*Figure 1:* (a) XRD of both bulk and 2D Cr₂Te₃, (b) AFM image showing flake distribution sizes (inset: lateral thickness distribution of the exfoliated flakes), (c) HAADF-STEM showing atomic arrangements along with FFT pattern (inset) of a single flake, where the hexagonal pattern can be seen together with the lattice spacing, (d) Raman spectra of 2D Cr₂Te₃, (e) M-H loop at temperatures ranging from 10 K to 300 K (50 K intervals), (f) Change in coercivity with respect to temperature (g) Cr₂Te₃ FC-ZFC M-T curve at 1000 Oe, and (h) dM/dT vs. T curves showing $T_C$.

**Figure 1d** shows the Raman spectrum of the exfoliated sample in the range of 50 to 200 cm$^{-1}$ at RT. The Raman spectrum was obtained using a commercial WITec Alpha 300 confocal Raman microscope. An excitation light source of 532 nm (Nd:YAG laser) with a spot diameter of 1 µm was used to probe the samples. The charecteristics peaks are mapped at 119.56 and 137.07 cm$^{-1}$ at room temperature, for the exfoliated structure. The peaks are similar to polarized raman peaks obtained for Cr$_2$Te$_3$ [41]. We obseve a blueshift in the E$_g$ peak which is a characteristic of coulombic interaction and variations in interlayer bondings [42]. The decreased number of layers and the different phonon modes affect the Raman shift. Raman active modes A$_1$ and E$_g$ have even parity under inversion. This is in close agreement with similar monolayer studies of Cr$_2$Te$_3$ [42–44]. Due to the convolution of many Raman active modes which highly depend on various system configurations, the major peak positions are affected. On prolonged laser exposure due to high thermal intensities, degradation of Te is usually observed, a similar trend observed in other telluride compounds [10,33,34,45,46]. Te is known to have large scattering cross section and even with the smallest penetration depth of the layers on these thin Te flakes results in surface precipitate formation [47].

MAGNETIC BEHAVIORAL STUDIES

Cr$_2$Te$_3$ has undergone substantial study for its magnetic behavior changes in lower dimensions [48–50]. Magnetism originates due to the ordering of the magnetic moment, which is localized at Cr sites and causes half-metallicity in Cr compounds [51]. Bulk Cr$_2$Te$_3$ are metallic ferromagnets [52], which was also observed from *M-H* curves of Bulk sample at room temperature (**Figure S4 a**). The theoretically calculated values of Te's induced magnetic polarization are antiparallel to the Cr localized moment, with average computed values of Te = 0.18 µ$_B$/Te and Cr = 3.30 µ$_B$/Cr [52]. Magnetic measurements were performed using a superconducting quantum interference device (SQUID) to determine the magnetic characteristics of 2D $Cr_2Te_3$ in comparison to its bulk counterpart. At room temperature, the 2D $Cr_2Te_3$ sample exhibited paramagnetic (PM) behavior as compared to its bulk parent material (**Figure S4 a**). In **Figure 1e**, *M-H* curves for 2D $Cr_2Te_3$ at various temperatures ranging from 10 K to 300 K (at intervals of 50 K) were plotted with an applied field between −10 and 10 kOe. The sample exhibits strong ferromagnetic (FM) behavior at lower temperatures below 250 K, and as it reaches RT, it exhibits paramagnetic behavior. The saturation magnetization of the 2D sample at RT was not attained even at magnetic fields (MT) of 60 kOe (6 T), justifying paramagnetic behavior as seen in **Figure S4 b**. The broadening of hysteresis was observed in the sample as the temperature decreased. The spontaneous symmetry breaking below $T_C$ allows the material to retain the net magnetic moment even in the absence of magnetic fields. The FM nature of the material is observed when the Cr - Cr distance in the lattice increases. This phenomenon appears due to strain in the material introduced during exfoliation. Therefore, strain tuning in the material resulted in a stable structure showing intrinsic magnetism and FM ordering at lower temperatures (10 K). In the absence of an external field, the sample still possesses a positive net magnetic moment and susceptibility. Magneto-crystalline anisotropy due to reduced crystal symmetry promotes long-range FM ordering in the 2D system, which is consistent with the first principles calculations. From **Figure 1f**, a high coercivity of 703.68 Oe was observed at 10 K. Few studies pertaining to monolayer $Cr_2Te_3$ show band structure change around $T_C$, which implies local magnetic moment persisting in the paramagnetic state [53]. There is a sharp increase in the magnetic moment ($\chi$), as moments align with the externally applied field.

Values were recorded during cooling in the 1000 Oe field, as we obtained *FC-ZFC* data for both samples at zero field cooling (*ZFC*) from 350 to 10 K. Following temperature stabilization, measurements were collected while the sample warmed up using a magnetic field of 1000 Oe.

From **Figure 1g,** the *FC-ZFC* data shows a Curie temperature ($T_C$) defined as the peak in the derivative. dM/dT vs. T (K) was performed to observe the transition temperature, as seen in the **Figure 1h** at ~290 K during FC and ~295 and during ZFC. From the above figures the 2D sample shows $T_C$ in the range of ~ 290 K - 295 K, where we observe a ferromagnetic to paramagnetic transition in the 2D material. To compare with the bulk parent sample, we measured the same amount of sample weight in SQUID at 1000 Oe, and the sample exhibited $T_C$ at ~180 K (**Figure S4 c**) similar to reported values [49,50]. The theoretically predicted values were at ~ 165 K which are due to contributions from Cr atoms. The theoretical values vary as we consider structures without defects or vacancies that are usually created during exfoliation of the sample. In order to confirm the magnetic phase transition. We perform differential scanning calorimetric measurements (DSC) on 2D $Cr_2Te_3$ as seen in **Figure S4 d**. The magnetic transition is a second order transition, to quantify the peaks we find the derivative dH/dT vs. T(K). As shown in **Figure S4 e**, the heat flow curve determines that the transition from ferromagnetic to paramagnetic occurs from 286.3 K (heating) and the same transition is observed at 286.9 K (cooling). Above $T_C$ temperature, a weak dipole interaction exists until room temperature, which is the origin of the spin magnetic moment induced in the 2D $Cr_2Te_3$ as observed from theoretical calculations as well.

RAMAN SPECTRA ANALYSIS

Reversal magnetism was observed in 2D $Cr_2Te_3$, with ease of structure reorientation upon application of an external magnetic field. The structural phase transitions were thoroughly studied using surface-sensitive Raman spectroscopic techniques. Both low-temperature and magnetic field effects on the 2D $Cr_2Te_3$ flakes were observed, as shown in the schematic of **Figure 2a**. For the low-temperature conditions, temperatures ranging from 0° to 24° (near room temperature) were considered. The experimental setup involved a silicon substrate on which the sample (2D $Cr_2Te_3$ flakes) was drop cast and placed on a cold plate (ice pack). The temperature was monitored using a handheld pyrometer. Measurements were taken at steps of 3° changes in temperature. The gradual change in temperature was attained by heating the sample with hot air from a dryer. Raman spectra for different temperature conditions are shown in **Figure 2b**. The major peaks are observed at 119.5 cm$^{-1}$ and 137 cm$^{-1}$ mapped for $A_1$ and $E_g$ peaks [15]. A new peak formation is observed at 175 cm$^{-1}$ along with the two main Raman bands. As the temperature increases, the peak intensity was more prominent. The ratio of peak intensity for $A_1$ and $E_g$ peaks are mapped in **SI Figure S5**. There is a decrease in $A_1$ intensity and an increase in the $E_g$'s. This shows heating effects in additional interlayer interactions other than van der Waals forces, as seen in other monolayer TMDs [54]. The intensity ratio does not follow a linear trend, as observed in **Figure S5**; it increases over a certain period and decreases as the temperature increases. Along with this, we observed the appearance of new peaks at 418, 644, and 680 cm$^{-1}$ (**Figure 2c**). These peaks are attributed to the formation of $Cr_2O_3$ (Chromia) [55,56], and the intensity of the peak increases with increasing temperature. The intensity variation in the spectra reveals oxide growth. This peak formation appears due to metal oxide bond formation upon constant application of laser heat, which in turn induces mechanical strain in the sample. In **Figure 2b,** we observe a left shift (redshift) in the Raman peaks by 5.53 cm$^{-1}$ for the $A_1$ peak and 1.25 cm$^{-1}$ for the $E_g$ peak, respectively. This redshift means that there is a lattice expansion during heating that is observed in the sample, which results in vibrational modes losing energy during the transition. Previous studies show that the strain dependence of the $A_1$ peak intensity exponentially decreases with isotropic tensile strain [57], but there is no decrease in intensity for uniaxial strains. The lattice strain effect is due to the heating of the sample. The trend in the peak shift remained the same during the heating and cooling of the sample substrate (**Figure 2d** and **2e**). This shows that the structure attains its original state and can be used during multiple heating and cooling operations. The thermal strain in the material was perfectly reversible; as the temperature increases, a decrease in wavenumber is observed [56]. The material shows reversible structural transformations, as observed during the heating and cooling of the sample.

On a similar trend, we investigated the structural changes in the 2D $Cr_2Te_3$ flakes when exposed to constant magnetic fields (neodymium magnets) over a time period. This was done in order to study the effects of an externally applied magnetic field exposure for deicing (discussed in the subsequent section). Upon constant magnetic field exposure, Raman shifts in peaks are observed for the peak $A_1$ from 119.4 to 123.28 cm$^{-1}$ (3.88 cm$^{-1}$), within 60 mins of magnetic field exposure (**Figure 2f**). The blue shift in the peaks can be attributed to interfacial strain effects between the layers. The $E_g$ peak shows a slight shift of approximately 1.15 cm$^{-1}$. The $A_1$ intensity does not change much for uniaxial strains but shows some shift because of lattice tensile strain in the material due to laser heating. However, after 90 min, the peaks start to disappear over time and shift to a larger wavenumber for both peaks, which is not desirable because the structure changes completely. Therefore, we maintain the external magnetic field exposure for a time period of no more than an hour for applications. We show that the structural properties of 2D $Cr_2Te_3$ can be reversed within this time frame.

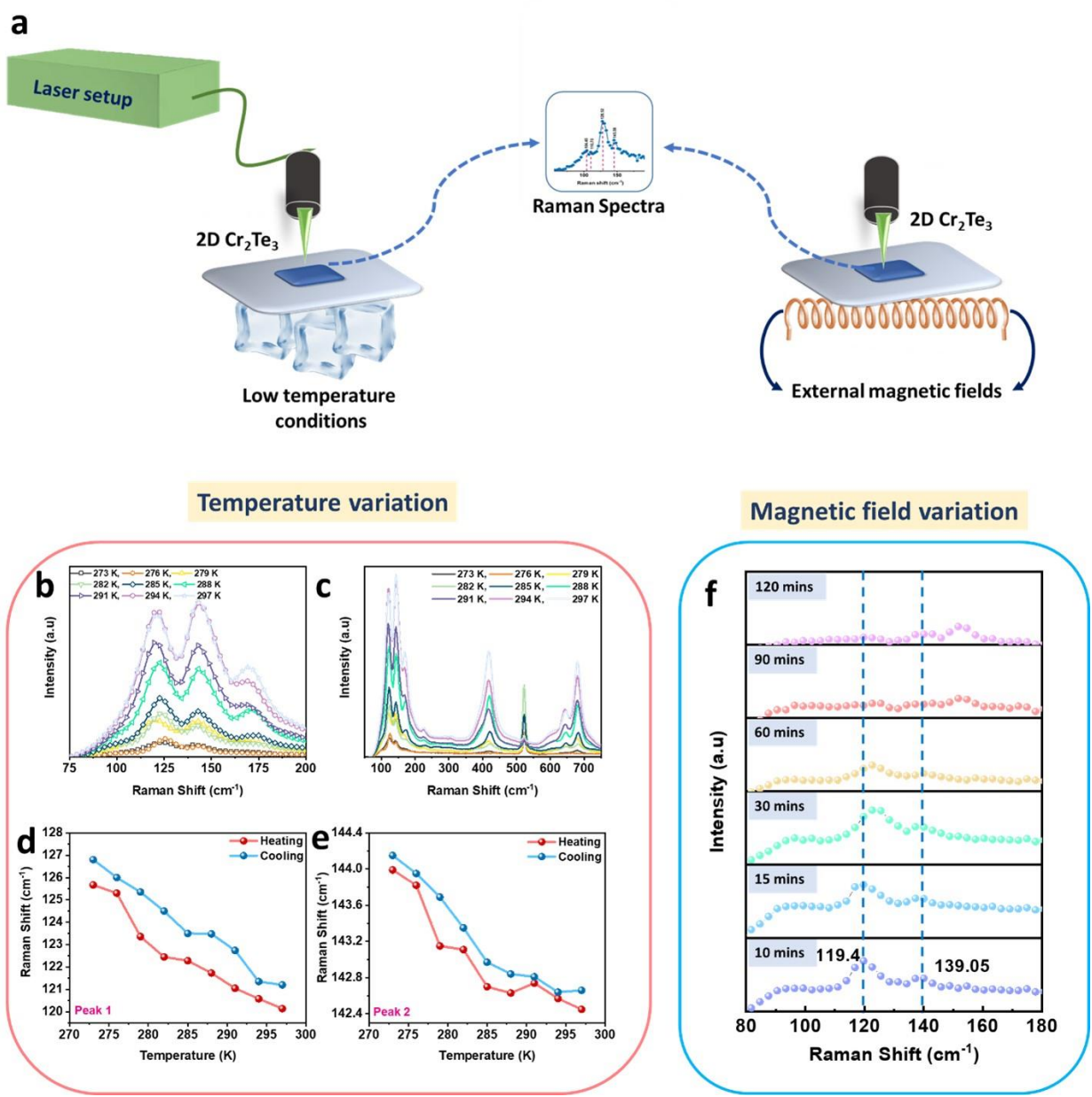

*Figure 2:* *(a) Schematic representation of the in situ Raman spectroscopic setup, (b) major 2D Cr₂Te₃ peaks mapped at different temperature conditions, (c) Raman spectrum showing peak formations at various regions; Main Raman band intensity variation during heating and cooling of the sample, (d) Peak 1 at 120 cm⁻¹, and (e) Peak 2 at 142.4 cm⁻¹, and (f) Raman spectra at varied magnetic fields.*

MODELING

Before proceeding to the theoretical numerical study of the electronic and magnetic structure of the material, it is necessary to carry out the geometry optimization of the proposed structural model, i.e., determining the optimized unit cell parameters (crystal vectors and angle values) and the atomic coordinates of the atoms in the unit cell, which is based on the principle of the minimum total energy of the system at a given external pressure [31,32]. Additionally, in the case of obtaining a spin-polarized solution, it is necessary to establish the ground state for the magnetic subsystem. In this paper, the preparation of a quasi-two-dimensional structure for its numerical modeling was carried out on the basis of the volume phase $Cr_2Te_3$ in the visualization program VESTA and the subsequent relaxation of the resulting system using the software package VASP 5.4.4, which is based on density functional theory (DFT) [58]. (**Details in Materials and Methods section**).

The final relaxed 2D structure of $Cr_2Te_3$ was investigated (**Figure S6a**, **Figure S6b**) considering the magnetic sub-lattice of chromium possessing an FM order. In order to prove the FM ordering of ground state, the total energies of AFM states have been calculated based on the unit cell containing 8 Cr-atoms: there are 35 possible AFM combinations of Cr spin orientations, and we have tested the 8 most probable (**Figure S7 (a-h)**). The corresponding free energies for each investigated antiferromagnetic structure are presented in **Table 1**. The ferromagnetic configuration is found to have the lowest energy, which proves the ferromagnetic nature of the 2D phase. Hence, within the framework of the mean-field approximation, the Curie temperature can estimated for 2D $Cr_2Te_3$:

$$k_B T_C \approx \frac{2}{3} S^2 \tilde{J} \approx \frac{2}{3} \frac{(E^{AFM} - E^{FM})}{2 \cdot N \cdot z} \qquad (2)$$

where $E_{AFM}$ and $E_{FM}$ are total energies per unit cell of ferro- and anti-ferromagnetically ordered system, $\tilde{J}$ is the average isotropic Heisenberg exchange parameter in terms of the first nearest-neighbor approximation, $k_B$ is the Boltzmann constant, N is the number of atoms in the cell and z is a specific coefficient related to the sublattice type and number of nearest neighbors in it, respectively, S is the spin moment.

Taking the lowest energy difference between the FM and AFM configurations (**Table 1**), $N = 8$ and $z \simeq 3$, we obtained $T_C \approx 165K$. Despite the fact that mean-field approximation generally tends to overestimate critical temperature, as experimentally observed position of curve slope was in the range of 285 K– 295 K (**Figure 1g**). The calculated cleavage energy for $Cr_2Te_3$ is found to be 0.64 J/m$^2$, which is in the same order of $MoS_2$ exfoliation energy[59]. This corresponds to experimental observations where $Cr_2Te_3$ was exfoliated into separate flakes by liquid phase exfoliation technique. Calculated values for lattice parameters are in good agreement with the experimentally measured lattice spacing (**Table 2**).

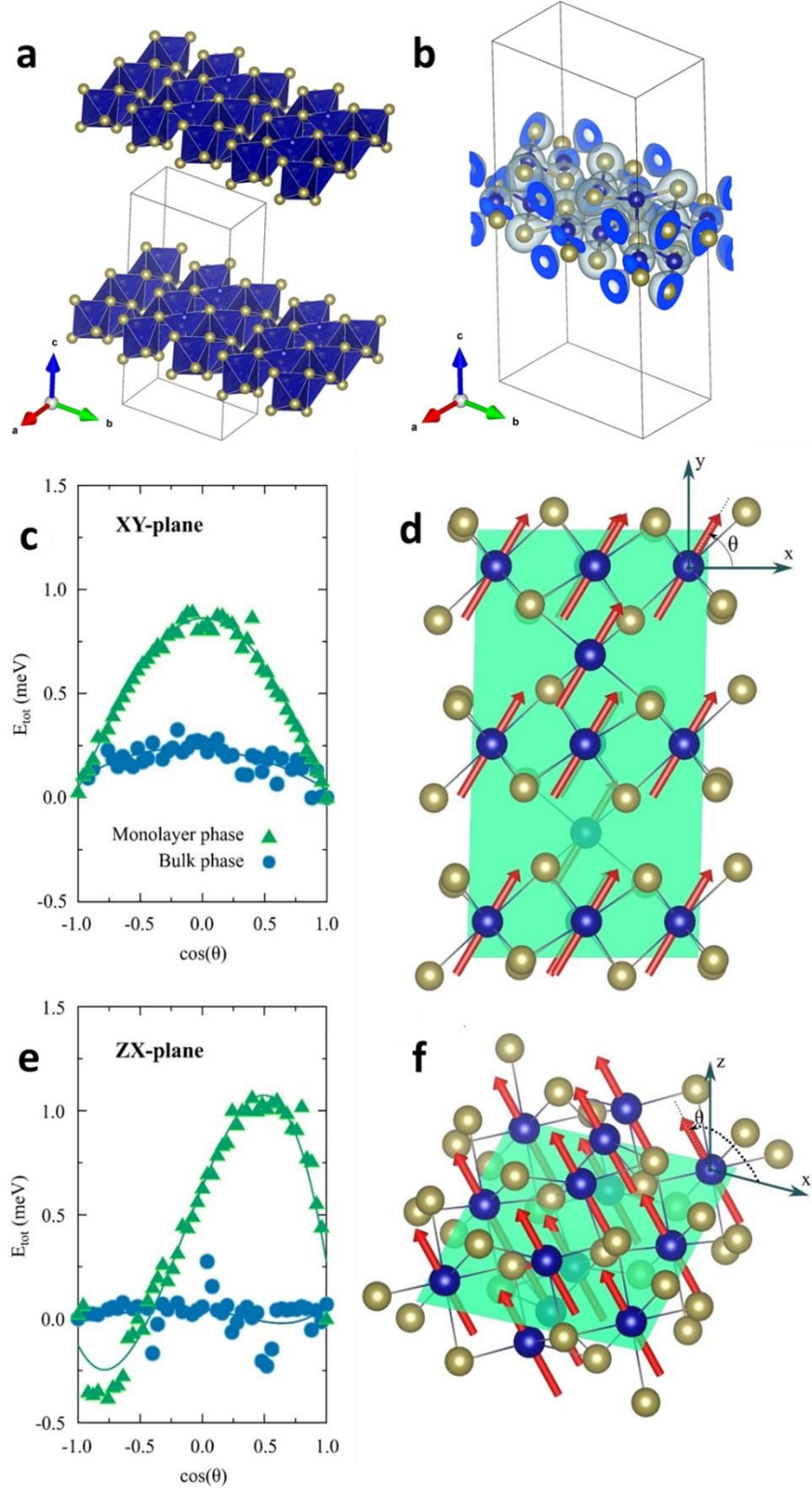

***Figure 3.*** *(a) Crystal structure (Cr-centered polyhedra view) and (b) charge density distribution for 1 L-$Cr_2Te_3$ (e/Å$^3$ isosurface), (c, e) Total energy per unit cell as a function of the angle cosine*

*between the Cr magnetic moments and the x-/z-axis placed in the xy-/zx-plane for monolayer and bulk phases, respectively; (d, f) Schematic representation of Cr magnetic moments orientations placed in the xy-plane and in the zx-plane; the xy-plane corresponds to the (1T̄0) crystallographic plane in the bulk phase while z-axis is perpendicular to the monolayer plane. Cr and Te atoms are depicted in blue and yellow, respectively; Lines correspond to the fourth-order polynomial interpolation. The zero energies for (c) and (e) inserts correspond to the ferromagnetic configurations with Cr spins aligned along the x- and z-axis, respectively.*

A non-collinear DFT formalism incorporating spin-orbit coupling (SOC) was used to estimate the anisotropy of bulk and monolayer $Cr_2Te_3$ phases in order to further investigate their magnetic behavior. The total energy of the unit cell is displayed in **Figure 3(c, e)** as a function of the angle between the Cr magnetic moments and the x-/z-axis placed in the xy-/zx-planes. The **Figure 3(d, f)** inserts illustrate the spin orientation of the Cr-sublattice in the xy-/zx-planes (induced magnetic moments of Te atoms were neglected). In contrast to the monolayer system, the anisotropy energy per formula unit of the bulk $Cr_2Te_3$ sample was found to be substantially lower, in the order of 0.06 meV/$Cr_2Te_3$ and 0.28 meV/$Cr_2Te_3$, in the xy- and zx- planes, respectively. Moreover, considering the strong FM nature of the Cr-spin interaction and the shape of the $E_{tot}\Theta$ curve having maxima at $\Theta \approx 60°$, such a complex exchange interaction can induce a nontrivial spin-canting and long-range ferromagnetic ordering. This shows that the atomic quasi-2D ordering significantly contributes to the enhancement of anisotropy energy. Furthermore, the predicted spin-canted long-range FM ordering indicates the potential for unique magnetic states in this system.

## MAGNETIC DEICING – PROPOSED CONCEPT

From safety and operational perspective, icing is a major cause for concern in aircrafts and other automobile. We tested 2D $Cr_2Te_3$ both as a dispersion and as a coating with a polymer. A single drop of the solvent with 2D $Cr_2Te_3$ was used for drop testing from a certain height in order to study droplet stabilization on the iced surface (**Supporting Information Video 1**). The solvent droplet movement in air and on the surface of the substrate is shown in the schematic in **Figure S8a**. A prototype condition of the iced surface was made by freezing a beaker using liquid nitrogen, as shown in **Figure S8b**, and a small section is enlarged in **Figure S8c**. Initially, blank solvent (IPA) was dropped on the edge of the beaker. As seen in **Figure S8d**, we observe only a small portion of the surface being covered. Later, 2D $Cr_2Te_3$ dispersed in IPA was dropped on the iced surface, and this was observed upon the application of an external magnetic field (Neodymium magnets were placed inside the beaker). On application of magnetic field, a single drop was able to cover a larger surface area (**Figure S8 e-j)**, which shows that the material can be dissolved and sprayed on iced surfaces, which can be directional. The droplet spreading took only a few minutes to cover a large area of the frozen surface. Concentrated magnetic fields at lower temperatures aids ferromagnetic $Cr_2Te_3$ in dissolving the iced surface. To assess the interfacial interactions and thermodynamic stability of the material, it is crucial to determine the surface free energy components and formation energies of the magnetic material [60].

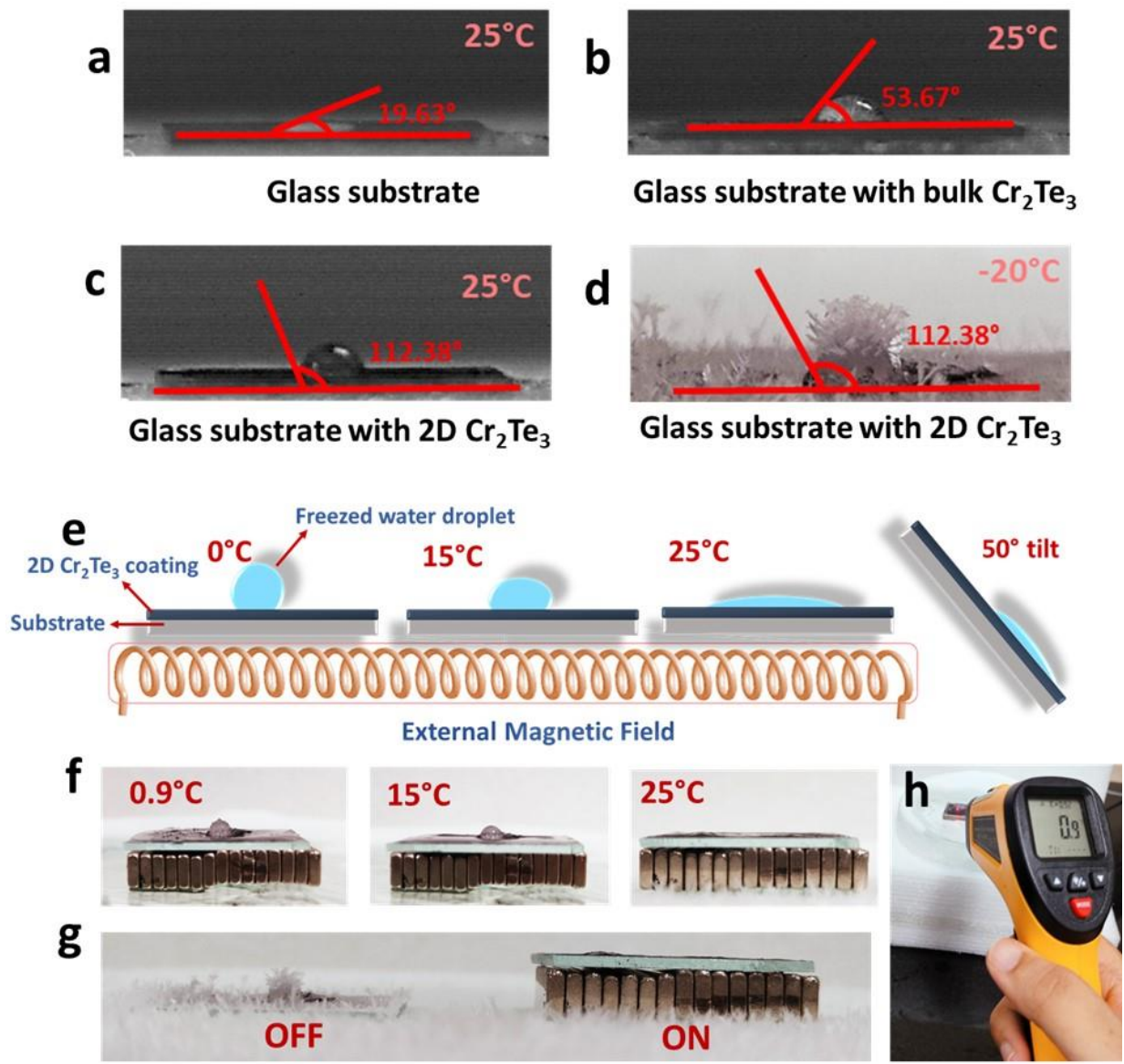

*Figure 4:* Contact angle measurements of water droplets on various surfaces at room temperature (a) plane glass substrate, (b) bulk $Cr_2Te_3$ coating, (c) 2D $Cr_2Te_3$ coating, (d) 2D $Cr_2Te_3$ at - 20°C, (e) Schematic representation of drop melting on glass substrate with 2D $Cr_2Te_3$ coating under external magnetic fields (ON conditions), (f) Pictographic representation of droplet melting over a time span of 10 secs, (g) Droplet on glass substrate with and without external magnetic fields and (h) Temperatures were monitored using a handheld pyrometer (0.9 °C temperature of the frozen droplet).

In the next set of experiments, we used 2D $Cr_2Te_3$ as a coating instead of a dispersant. Deicing requires hydrophobic surfaces to repel water droplets condensed on the surface and/or ice crystals. Therefore, surface studies, such as contact angle measurements, were performed to study its hydrophobic/hydrophilic nature. The surface interaction was further investigated with measurements of the water contact angle (WCA) on the $Cr_2Te_3$-coated glass substrate. The coating consists of a polymer matrix with dispersed 2D $Cr_2Te_3$ flakes. A polymer is usually used to disperse the magnetic sample; this acts as a matrix for the particles and helps in binding the material on other surfaces[61]. Magnetic samples prefer to disperse or aggregate by tending to attract each other depending on the media. Thus, the polymer prevents agglomerations and oxidation during coating[62]. Polyvinylidene fluoride (PVDF) was dispersed in acetone (concentration - 10 mg/ml). The sample-to-polymer solvent ratio was kept at 0.5 mg: 10 ml. Glass substrate was used to study the droplet effect mechanism on the windshields, windows of aircraft, and other vehicles in colder regions. WCA was measured using a Fastcam Mini UX50 camera and analyzed using Photron photo viewer software (PFV4 (×64)) (Specifications in **Supplementary Information** and **Figure S9 and S10**). Initial measurements were done at room temperature. **Figure 4a** shows a water droplet on a blank glass substrate, and the measured WCA was 19.63°. This shows the hydrophilic nature of the glass substrate, where the residue remains after a longer duration. To achieve lower surface energy and low polarizability, electronegative materials are generally preferred. Later, the WCA of the bulk $Cr_2Te_3$ + polymer-coated surface was measured to be 53.67° (**Figure 4b**), which was better than that of plane glass but still hydrophilic in nature. Changing the coating to a single layer of 2D $Cr_2Te_3$ polymer composite shows an increase in WCA to 112.38°, which acts as a hydrophobic surface (> 90°), as seen in **Figure 4c**. The droplet was observed for a time duration of >120 mins, and the WCA remained stable. **Figure 4d** shows a frozen droplet on top of 2D $Cr_2Te_3$ surface with WCA at -20 °C remains the same at ~112°. Multiple layer coatings resulted in overlapping PVDF layers showing non-adhesive layering, and the $Cr_2Te_3$ flakes started to shed along with the polymer layer. Therefore, one layer of the polymer composite along with 2D $Cr_2Te_3$ material (optimum) was used for better hydrophobic property exploration.

In **Figure 4e**, we present a schematic representation of the 2D $Cr_2Te_3$ coated glass substrate with external MFs in the ON condition. This shows that once the magnetic coils are ON, the icing on the surface starts to melt instantly. **Figure 4f** shows the ice melting at different temperatures (0°C to 25°C) just with the application of a magnetic field, and no heat was provided or generated during the process. This temperature variation causes changes in the magnetic properties of 2D $Cr_2Te_3$. **Figure 4g** shows a clear demarcation of OFF and ON conditions without and with external magnetic sources, respectively (**Supplementary Video 2**). We try to observe the behavior of the 2D $Cr_2Te_3$ material in extremely cold conditions, with and without external magnetic field effects.

DISCUSSIONS

(1) THERMODYNAMIC POINT OF VIEW

One of the possibilities of the observed phenomenon may be due to the structural change in the material which induces changes in the atomic vibrations, which influence magnetic moments. This can be expressed in thermodynamic terms as magnetic materials can be thought of as a system composed of three main energy reservoirs, mentioned in the equation below. The total entropy of the system at constant pressure can be expressed as:

$$S_{Total(T, H)} = S_{L(T, H)} + S_{M(T, H)} + S_{E(T, H)} \qquad (1)$$

where $S_{L(T,H)}$ is the entropy contribution by vibrational excitation of the lattice, $S_{M(T,H)}$ is the entropy contribution by magnetic states of the magnetic sublattice, and $S_{E(T,H)}$ is the entropy contribution by electronic configuration associated with electronic bands. When an external magnetic field is applied, the magnetic moments reorientate themselves, leading to a change in entropy associated with magnetic sublattice ($S_{M(T,H)}$). When the process mentioned above occurs under adiabatic condition ($\Delta S = 0$), changes in the magnetic entropy can be compensated by changes in lattice entropy ($S_{L(T,H)}$) in order to maintain the total entropy ($S_{Total(T,H)}$) constant. The energy transfer taking place from the magnetic reservoir to the lattice one results in a small temperature increase, which aids in the deicing of frosted surfaces (magnetic deicing), one of the application that we have explored further in this work.

Also, the minimization of the surface Gibbs energy, $\Delta G = \Delta H - T\Delta S$, where H, T, and S are the enthalpy, temperature, and entropy, respectively, drives hydrophobic contacts and ice formation on top of the aircraft surface. This is because a large positive value of $T\Delta S$ in hydrophobic contacts outweighs a modest positive value of $\Delta H$, making spontaneous hydrophobic interactions energetic. At the critical temperature *Tc*, where the entropic contribution to the Gibbs energy, $T\Delta S$, prevails over the enthalpic contribution, $T\Delta H$, makes it more energetically advantageous for the ice crystal to be rough, the surface roughening transition governs the direction of ice crystal growth. This shows that icephobic and hydrophobic behaviors can both be understood as entropic effects from a thermodynamic perspective. To understand this phenomenon we study the surface effects of the $H_2O$ molecule with $Cr_2Te_3$ theoretically in the next section.

(2) SURFACE INTERACTION POINT OF VIEW

A wide range of research exists on water molecule interaction with magnetic particles. Some of them include static magnetic fields changing the hydration level of the shells[63], increased pH levels[64], and aiding water evaporation[65]. Ferromagnetic materials have the ability to switch atomic dipoles. In water molecules, hydrogen has a positive charge, and oxygen has negative. Switching of surface properties during the application of a magnetic field was observed in the schematic shown in **Figure 5a**. Without magnetic fields condition (**Figure 5a**), the ice droplet does not interact with the 2D coating. On application of a magnetic field (**Figure 5b**), the spins reorient parallel to the *c-axis*, middle plane, or canted FM ordering, and the weak hydrogen bonds between the $H^+$ and $O^-$ ions experience partial molecular alignment. This leads to residual anisotropic magnetic interaction between the $H_2O$ molecule and ferromagnetic 2D $Cr_2Te_3$ material. **Figure 5c** shows the +/- charge possessed by each element without magnetic field condition. In **Figure 5d** with magnetic field condition, when ice droplet comes in contact with the 2D $Cr_2Te_3$ coating and experiences switching due to *c-axis* FM ordering (represented by arrows) in the material. The 2D $Cr_2Te_3$ provides a larger surface area, exposing $Cr^+$ ions on the surface for magnetic dipolar interaction with $H^+$ ions, which aids repulsion. The strong FM behavior at lower temperatures (< 0 °C) inhibits ice formation on the surface of the coating, therefore condensing the droplets formed on the surface on external magnetic field application.

During the surface interaction, we considered two cases (**Figure S12**), oxygen ions (O – oriented) reoriented towards ML $Cr_2Te_3$ surface and hydrogen ions reoriented (H – oriented) to the ML surface. The resulting 1L-$H_2O$/$Cr_2Te_3$ structures with oxygen ions oriented towards the chromium telluride plane (O-orientation) (**Figure S12a**) in the ferromagnetic and paramagnetic states are energetically preferable to the similar structures with H-orientation (**Figure S12b**). The formation of an O-oriented 1L-$H_2O$/$Cr_2Te_3$ structure in the ferromagnetic state was found to be about 60 % more energetically favorable than in the paramagnetic state (**Table S1**). The calculated vdW gap values between the ice and chromium telluride planes are 1.9/2.3 Å for the ferromagnetic/paramagnetic state. These values correspond to the hydrophobic and hydrophilic nature of 1L-$Cr_2Te_3$ in the ferromagnetic and paramagnetic states, respectively. For the magnetic case, we found the charge transfer as the difference between the charge densities of the 1L-$H_2O$/$Cr_2Te_3$ heterostructure and the bare ice and $Cr_2Te_3$ MLs with positive and negative charge densities as seen in **Figure 5e**. Experimental and theoretical results showed increased magnetic anisotropy and lesser formation energy for long-range ferromagnetic magnetic ordering in the 2D $Cr_2Te_3$. Thus, other FM materials can also be explored for low temperature (less than room temperature) deicing applications in areas such as automobiles, aircraft, and other architectures. However, the diamagnetic nature of water must also be taken into account, which means that when subjected to a magnetic field, it tends to repel the magnetic lines. We carried out the experiments under the same conditions using the non-diamagnetic fluid, such as coconut oil, and the results were much less pronounced (**Figure S14**). We were able to observe the saturated fatty oil freeze quickly as soon as it was dropped on the frozen surface (without external magnetic field), as seen in **Figure S14a**. The droplet remains frozen even at room temperatures (**Figure S14b**). The droplet was then exposed to same magnetic field strength as frozen water droplet, but the oil droplet does not repel of the surface. The oil droplet adheres strongly to the frozen glass surface (**Figure S14c**) and does not melt even when temperatures reached slightly above room temperature (> 25 °C) as seen in **Figure S14d**. This shows that non-diamagnetic liquids such as oils and other solvents do not exhibit the same property as that of frozen $H_2O$ molecule interacting with 2D $Cr_2Te_3$ surface. Investigating the effects of external magnetic field on the defect densities and vacancies created in 2D structures is an important aspect were we still lack clear understanding of the 2D material's behavior.

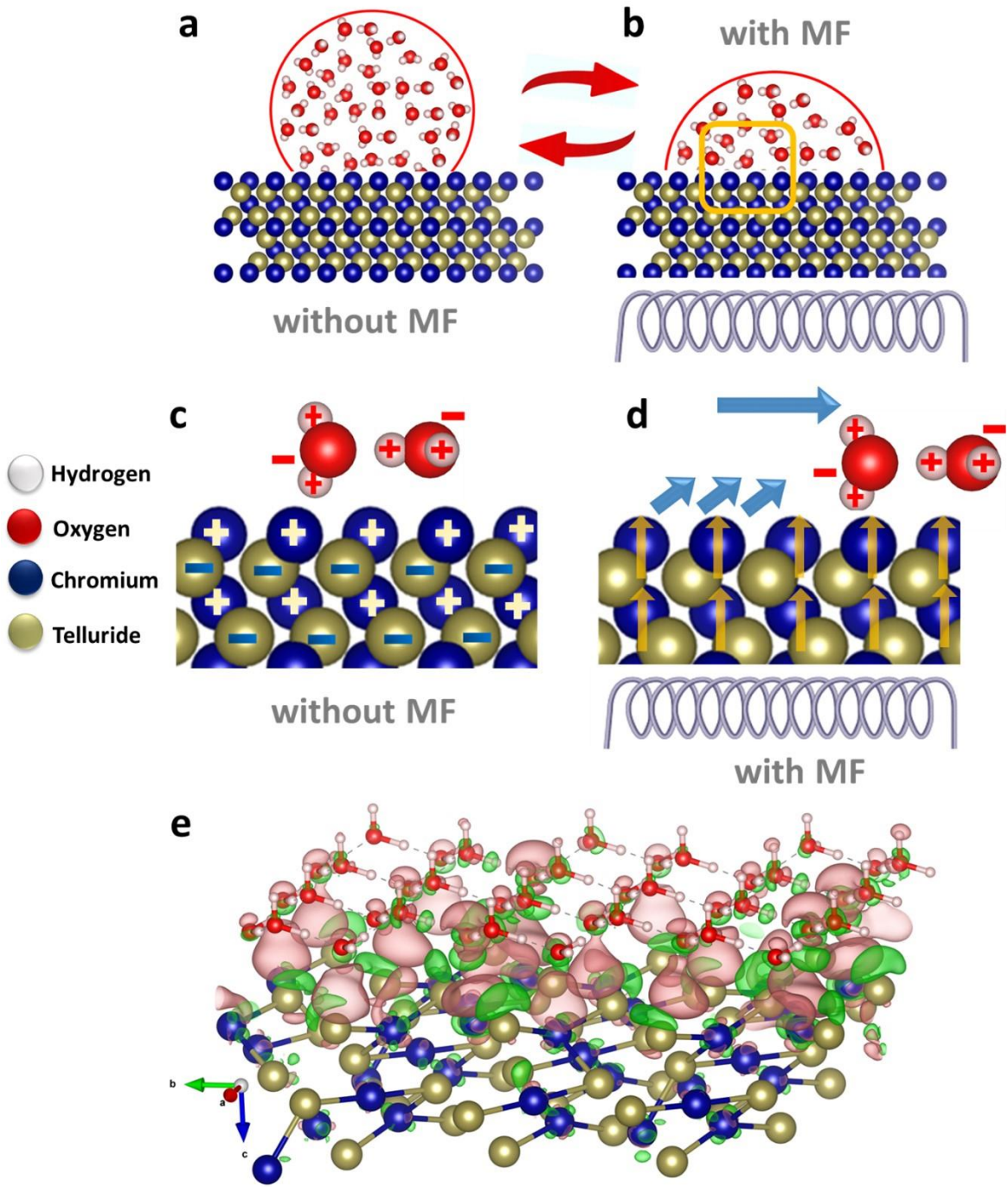

***Figure 5:*** *Ice droplet on 2D $Cr_2Te_3$ coating (a) OFF and (b) ON condition (external magnetic fields), (c) Charge interactions in OFF condition and (d) c-axis FM ordering in 2D $Cr_2Te_3$ on application of magnetic fields and repulsion of water molecule at low temperature (0 °C), (e) Charge transfer isosurface ($3 \cdot 10^{-4}$ e/Å) between ice and ferromagnetic $Cr_2Te_3$ monolayers. Red/green color corresponds to the negative/positive charge difference.*

CONCLUSIONS

Liquid phase-exfoliated 2D $Cr_2Te_3$ flakes exhibited size tunable ferromagnetic (FM) nature at lower temperatures and paramagnetic (PM) behavior approaching to room temperature. The 2D nature of the material was thoroughly characterized for its structural and magnetic behavior. We theoretically show that the structure cleavage plane of 0.64 J/m$^2$, is favorable for the spin magnetic moment induced in the exfoliated material. DFT calculations show that the monolayer of the structure influences long-range FM ordering, along with magnetic anisotropy. The critical transition temperature $T_C$ was observed in the range of ~285 – 295 K. This tunable FM nature of 2D $Cr_2Te_3$ was used to our advantage to study the deicing property of the material, along with the deicing property exploration of 2D $Cr_2Te_3$ embedded in a polymer matrix. The structural phase transition in the material was probed via *in-situ* Raman spectroscopic studies at various temperatures and magnetic conditions. The 2D material exhibits reversible crystal structure transformation during heating and cooling conditions (0° until 25 °C). With minimum formation energy ($E^{form}$) during surface interaction between $H_2O$ molecule and 2D $Cr_2Te_3$, and no heat generation, we were able to defrost the surfaces due to strong FM nature of the 2D $Cr_2Te_3$ at lower temperatures (< 0 °C). Long-range FM ordering along with interfacial strain effects in the 2D material can be used to address the existing major issue of icing in aircraft and other automobiles via simple coating of 2D $Cr_2Te_3$ material.

**Table 1.** Free energies of antiferromagnetic structures relative to the ferromagnetic configuration per calculation cell (a-h). The total energy for the FM structure is chosen as a zero point.

| Structure | Free energy, (meV/$Cr_2Te_3$) | Structure | Free energy, (meV/$Cr_2Te_3$) |
|---|---|---|---|
| AFM I (**a**) | 389.4775 | AFM V (**e**) | 287.455 |
| AFM II (**b**) | 314.455 | AFM VI (**f**) | 338.3025 |
| AFM III (**c**) | 256.9775 | AFM VII (**g**) | 319.825 |
| AFM IV (**d**) | 314.455 | AFM VIII (**h**) | 323.8025 |

**Table 2.** Computed lattice parameters for bulk and monolayered $Cr_2Te_3$.

| Structural form | Calculated parameters | Experimental values |
|---|---|---|
| Bulk | a=6.76846 b=6.76846 c=12.22340<br>α=90.0 β=90.0 γ=120.0<br>Unit-cell volume = 484.956406 Å$^3$ | a=6.8230 b=6.8230<br>c=11.8000 α=90.0<br>β=90.0000 γ=120.0<br>Unit-cell volume = 475.73 Å$^3$ |
| Monolayer | a=6.36436 b=11.52200 c=23.30775<br>α=90.0 β=90.0 γ=88.5777 | - |

**Competing interests**

The authors declare no competing interests.

AUTHOR INFORMATION

**Corresponding Author**

**\*Corresponding Author:** Prof. Chandra Sekhar Tiwary (chandra.tiwary@metal.iitkgp.ac.in), Prof. Douglas S. Galvão (galvao@ifi.unicamp.br) and Alexey Kartsev (karec1@gmail.com)

**Author Contributions**

The manuscript was written through contributions of all authors. All authors have given approval to the final version of the manuscript.